\documentclass[a4paper,twoside]{article}

\usepackage{epsfig}
\usepackage{subcaption}
\usepackage{calc}
\usepackage{amssymb}
\usepackage{amstext}
\usepackage{amsmath}
\usepackage{amsthm}
\usepackage{multicol}
\usepackage{pslatex}
\usepackage{apalike}
\usepackage{SCITEPRESS}     

\title{Machine Teaching by Domain Experts: Towards More Humane, Inclusive, and Intelligent Machine Learning Systems}

\begin{document}

\author{\authorname{Claudio S. Pinhanez}
\affiliation{IBM Research Brazil, Rua Tutoia 1157, Sao Paulo, Brazil}
\email{csantosp@br.ibm.com}}

%
%

\abstract{
This paper argues that a possible way to escape from the limitations of current machine learning (ML) systems is to allow their development directly by domain experts without the mediation of ML experts. This could be accomplished by making ML systems interactively teachable using concepts, definitions, and similar high-level knowledge constructs. Pointing to the recent advances in \emph{machine teaching} technology, we list key technical challenges specific for such expert-centric ML systems, and suggest that they are more humane and possibly more intelligent than traditional ML systems in many domains. We then argue that ML systems could also benefit greatly from being built by a community of experts as much as open source software did, creating more inclusive systems, in terms of enabling different points-of-view about the same corpus of knowledge. Advantages of the community approach over current ways to build ML systems, as well as specific challenges this approach raises, are also discussed in the paper. }


\onecolumn \maketitle \normalsize \setcounter{footnote}{0} \vfill

\section{\uppercase{The Autumn of Deep Learning}}

We are arguably living in the autumn of \emph{Deep Learning (DL)}, when \emph{Artificial Intelligence (AI)} has gone from exploding in the news and minds as an revolutionary technology to a more sober reality where its limitations and drawbacks are becoming clearer. We are harvesting the benefits of a great summer, but we must prepare for  winter by exploring new ways of creating \emph{Machine Learning (ML)} systems.

We have indeed witnessed in the last years admirable intelligent machines which have challenged the boundaries of what academics considered possible. The \emph{IBM Watson} computer, based on probabilistic machine learning, beat the best players in the most difficult trivia contest in the world; the \emph{AlphaGo} program humiliated one of the best Go players in the world, just to be defeated by \emph{AlphaZero} some months later; ML datasets had their previous records shattered by deep learning programs; machines started to learn by playing against adversarial machines using \emph{GAN} techniques; self-driving cars started to roam the streets; and real-time speech recognition arrived at home, providing news, translation, and other language-related services. 

We are now almost close to a decade of such DL-related successes, and the momentum of such data-hungry ML technologies seems to be slowing down. First and foremost, because there are many domains and practical problems where getting data suitable for DL is impractical or impossible. Second, controls and regulations over the use of personal data have increased in recent years. Third, in many applications the opacity of DL systems makes them often not appropriate to be in key decision-making processes. Fourth, the use of AI to manipulate elections and media has created a lot of concern in the society around those technologies. Lastly, the possible impact of AI in jobs and in the economy has generated many calls for control and regulation. 

\begin{figure}[t!]
  \centering
  \includegraphics[width=7.5cm]{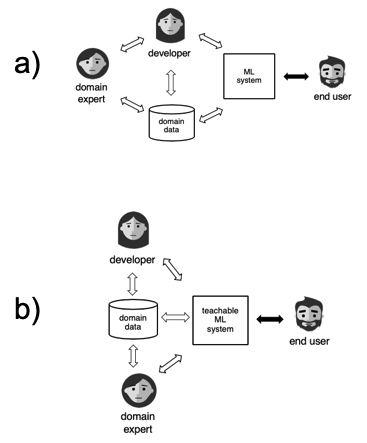}
  \caption{Different ways to create ML systems: a) mediating domain expert(s) through ML developers; b) direct access of domain experts using machine teaching.}
  \label{fig:typemlsystems}
\end{figure}

We thus believe AI is due to a new wave of ideas and approaches which can handle appropriately some of the issues listed before. This paper examines how some of those new approaches of machine learning, still in the early stages of development, generically referred here as \emph{Machine Teaching (MT)}, can lead to significant changes in the way ML systems are developed, and how they can lead to more humane, inclusive, and intelligent machines. In particular, we look into machine teaching technologies which may allow domain experts to have direct access to the construction of ML systems, which today is, in most cases, mediated by a ML-expert developer (see Figure~\ref{fig:typemlsystems}). We propose that re-focusing the centricity of the development process from ML-experts towards domain experts can lead to better ML systems in terms of improved intelligence.  We also believe machine teaching (and associated technologies) has the potential to empower domain experts to free them from the machine-like tasks of labeling data for machines. 

\begin{figure}[t!]
  \centering
  \includegraphics[width=5.8cm]{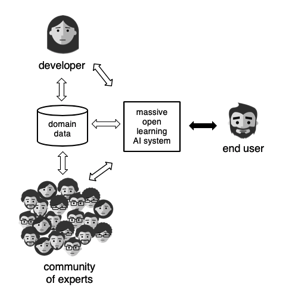}
  \caption{A massive open learning AI system (MOLA).}
  \label{fig:communitymlsystems}
\end{figure}

Moreover, directly contact to ML systems by domain experts may lead to the development of intelligent systems by communities of domain experts, in similar ways to open source software. We call a ML system which learns from a large, diverse, and mostly unmediated community a \emph{Massive Open Learning AI system}, or \emph{MOLA} (see Figure~\ref{fig:communitymlsystems}). A fundamental difference between a MOLA and a teachable ML system (Figure~\ref{fig:typemlsystems}.b) is that a MOLA can interact directly and simultaneously with a community of experts, with different kinds of knowledge and experience, different opinions and approaches to the subject matter, and different degrees of commitment to the system.

To simplify the terminology in this paper, we use the term \emph{machine} to refer to any computerized system which can be developed by human beings to perform some task. This can be accomplished either by traditional programming or by obscure training of neural network-inspired systems. We also use \emph{machine learning} as a generic term including probabilistic machine learning, graphic models, neural networks, reinforcement learning, and, of course, deep learning. We make the simplifying choice of using the term \emph{deep learning} to refer to systems which use numerical embedding architectures and learning algorithms which require massive amounts of labeled data to be trained, such as traditional \emph{Neural Networks}, \emph{CNNs}, \emph{LTSMs}, and similar technologies. Also, we use the term \emph{experts} to refer to professionals with expertise in a given domain, and distinguish it from \emph{ML-experts}, understood here as developers and researchers who are knowledgeable of machine learning techniques.

We start by looking into how data labeling frames knowledge transfer into an inefficient and inhumane task for people, and  often leads to errors. Following we start to explore the differences between machines which 
are trainable (using examples) to those which are \emph{teachable} (using concepts). We then map those concepts in the context of previous literature of \emph{machine teaching} and \emph{interactive machine learning}, and look at the technical challenges involved in allowing experts to directly construct them. Next we explore the challenges and advantages of building an intelligent system using a community of domain experts instead individual or a small group of experts, and describe a hypothetical example of a community of nutritionists working together to create an AI assistant for people with diabetes. We close by exploring the challenges and advantages of taking the community approach and discussing possible negative social impacts.

\section{\uppercase{Labeled Data Considered Harmful}}

Deep learning methods are deep-hungry for labeled data which can be easily mapped into the input and output of mathematical functions. 
Three great schemes have been used to obtain the labeled data: data created by machines (a.k.a. \emph{Internet of Things} data); ``appropriation'' of data which people have created for other purposes (such as social media data); or get if from human beings in ``labeling farms''.


In the context of this paper, we focus on the latter way to produce data in which
domain experts are provided examples of an input to the function to be constructed and asked for the output associated with them. The labeller can range from an anonymous worker using \emph{Amazon's Mechanical Turk} to a physician expert in rare cancers. In both cases, human beings are treated as machine feeders, working intensively to transform data into a form which can be easily digested by a ML algorithm. There is ample evidence that people do not perform well doing those tasks~\cite{aroyowelty2015,flexergrill2016}: they get easily bored, make frequent mistakes, and use concepts and rules whose semantics varies wildly with time~\cite{simard2017}. Also, a consistent and motivated labeller is rare to find.

In particular, high-level domain experts seem to hate to be in that position. This is a common problem faced by developers of ML systems for medicine and healthcare~\cite{gurarietal015}. Experts seem to dislike immensely being in the role of oracles. Also, studies have shown that when people are asked to provide labels to interactively train machines, they often try to use the feedback mechanisms provided for multiple, overlapping purposes~\cite{stumpf2009}. Experts easily drift into \emph{social learning}~\cite{thomazbreazeal2008}, trying to help and encourage the systems, and avoiding being too harsh to the machine.

It is an interesting paradox that in order to teach machines, experts have to behave like one of them. They have to label data consistently (often blindly to context),  mechanically and repetitively producing data which can be consumed by the training process of a machine. The richness of the human experience and knowledge is low-leveled so a machine can use it, paradoxically, to reproduce the complexity of human knowledge. Asking people to assign labels to data can be regarded as a form of dehumanization which is likely to be reflected in the quality of the generated data, as discussed by~\cite{blackwell2015}. But, more importantly, knowledge is bottle-necked through a simplistic input-output view of the world, and magically expected to reappear in its full glory in the machine behavior. Like if it was possible to grind beef to the point it can pass a small tube, and then pass it back through an ungrinding machine which re-creates it into a magnificent steak.

In a similar way, as pointed in~\cite{darwiche2018}, a lot of the DL recent success is also due to a great simplification of the original AI goals and tasks. For instance, many of the datasets in which DL performed extremely well either are very simple tasks for humans (such as distinguishing dogs from cats), or are considered in new contexts where criticality is low (see~\cite{darwiche2018}). A typical example are dialogue datasets in which researchers use success metrics which are almost devoid of semantics. 

Of course there are problems where the knowledge needed to solve them can survive this grinding-ungrinding process of reduction to input-output data. But it is naive to expect that this way to transfer knowledge can be generally applied to the solution of most of problems. We see that winter is coming for deep learning as a consequence of this simplistic, reductionist view of knowledge, and is likely to arrive in spite of all the engineering feats of its practitioners.


But before we move our arguments ahead, it is important to define more precisely what we mean by teaching a machine and, especially, to distinguish it from other forms of knowledge transfer. For us, \emph{teaching}, in the context of machines, is to transfer knowledge to a machine using elements of the process we normally employ to make human beings learn, using \emph{declarative} statements of \emph{concepts}, exemplars, definitions, demonstrations, procedures, and tests. 
We distinguish it from \emph{training}, which is when a ML-expert shows a machine \emph{examples} which define the outputs expected to given inputs, which are then combined to create \emph{inductive} patterns of behavior. 
\section{\uppercase{Expert-Centric Machine Teaching}}

In this paper we advocate the approach where (domain) experts, or a community of them, interactively teach, without significant mediation, a ML system. However, it is necessary to be more precise about what we denote as \emph{expert-centric machine teaching} since terms like \emph{machine teaching} and \emph{interactive machine learning} are quite overloaded. To do that, we start by reviewing the two main related areas of ML: interactive training and high-level teaching.

\subsection{Related Work}

The seminal work of Fails and Olsen~[2003],
which coined \emph{interactive machine learning}, looked into direct interaction of designers (experts) and the ML system but still viewed the experts' role as providing examples. Thomaz and Breazeal~[2008]
explored further this interaction by looking into how teachers appropriate feedback mechanisms to communicate different types of teaching acts, what was further explored by other researchers~\cite{fiebrinketal2011,kuleszaetal2015,hoodetal2015,senftetal2017,royetal2018}. There is evidence from previous studies that human beings prefer to teach machines~\cite{thomazbreazeal2008,kaocharetal2011,stumpf2009} than to train them, and also that they are better in the former than in the latter.  The dichotomy between the teaching and training has been a constant struggle in the field of AI, from the pioneering times of rule-based systems, logic programming, and some forms of probabilistic machine learning, to the engineering of neural networks of today.

Some works such as~\cite{zhu2015,liuetal2017,fanetal2018} have explored machine teaching mostly as a way to decrease the number of examples needed to train a system. \cite{amershi2014,dudleykristensson2018} are good surveys of user experience ideas and guidelines for interactive machine learning interfaces. 

Although tools for transferring knowledge from experts to traditional AI systems were explored in the past~\cite{quintanaetal2015}, the research of Stumpf \emph{et al.}~[2009] 
and Kaochar \emph{et al.}~[2011] 
were among the first to look into how experts could teach ML systems using concepts and methods beyond examples. In particular, \cite{fogartyetal2008,amershietal2015,simard2017} have explored machine teaching by interactive feature selection and composition. Recent works have looked into other aspects of machine teaching, including the teachers' mental models~\cite{sarkar2015}, textual description of procedures~\cite{azariaetal2016}, 
and natural language explanations~\cite{hancock2018}. 

Finally, the development of computer systems using large, often open, communities is also more common in the two extremes of the spectrum. Open source software has demonstrated that a large community can assemble and work together to program with high quality and reliability. Also, the use of crowd-sourcing in labelling examples for training, though a lesser form of community, is quite well succeeded. Teaching machines withing the context of a community is less common, and have resulted both in successes and failures. The \emph{NELL} project has succeeded in congregating a large community to build a system which is able to read, collecting more than 80 million confidence-weighted beliefs in the process\cite{mitchell2015}. On the other hand, Microsoft's \emph{Tay} has shown some of the difficulties such as when part of the community decided to sabotage the project by teaching the chatbot inappropriate behavior~\cite{neff2016}.

\subsection{Why Unmediated Experts?}

The AI community has faced before the need of AI-experts to translate into machine language the knowledge from a domain, in what is called \emph{Knowledge Engineering (KE)}. Although KE has evolved into a structured discipline~\cite{tecucietal2016}, in most domains the constraint of requiring non-experts to gather the related knowledge from an expert, and structure it for a machine, seems to have precluded the emergence of successful AI systems. 

On the other hand, some recent successes in enabling simplified direct access to simple ML systems show the promise of removing ML intermediates. For example, 
tool to create chatbots has a basic, intention-action interface, in which non-ML-experts can define basic intents using exemplar sentences and associate them to specific actions. The simplicity of the interface has fostered the creation of hundreds of thousands of chatbots worldwide by experts with very limited knowledge of ML. Although the intent-action model tends to make difficult the scale up of such efforts, thus limiting depth of knowledge and quality, this case has demonstrated that making the construction of an AI system available to end-user experts has a lot of potential. 

\subsection{Enabling Direct Access}

We argue here for exploring machine teaching techniques, existent and future, towards enabling experts' direct access to ML systems. We agree with~\cite{simard2017} that focusing on the teacher requires simplifying many tasks executed today by the ML experts, so it is important to identify the set of requirements for machine teaching technology which are specific to facilitate access to experts:
\begin{itemize}
    \item \textbf{Transparency:} domain experts tend to be more effective if they understand the inner workings of the system~\cite{thomazbreazeal2008}.
    \item \textbf{Explainability:} the ability of experts to teach an AI and its capability of explaining itself tend to be inter-related~\cite{leu2017}.
    \item \textbf{Multiple teaching patterns:} research has shown that human teachers employ several modalities of actions when trying to teach a machine~\cite{stumpf2009,thomazbreazeal2008}.
    \item \textbf{Fast re-training:} experiments~\cite{failsolsen2003} have shown that domain experts tend to be impatient (that is, less than 5 seconds) while waiting for ML systems to re-train itself. 
\end{itemize}

In practice, creating a machine teacher experience satisfying those requirements is a formidable challenge for both current ML and MT technologies. A typical DL system is hard to understand and explain, learns only from input-output examples, and re-training is often a matter of days in specialized hardware. Nevertheless, as discussed before, in all those areas there has been significant recent progress, which may be channeled and adapted towards creating expert-centric, teachable ML systems.

\section{\uppercase{From Groups of Experts to Communities}}

 By making direct interaction and transfer of knowledge from experts to ML systems possible, machine teaching techniques also open the possibility that a community of experts interacts and teaches a ML system, without mediation, about a particular subject or task. As mentioned, we call a ML system which learns from a large, diverse, and mostly unmediated community a \emph{Massive Open Learning AI system}, or \emph{MOLA} (see Figure~\ref{fig:communitymlsystems}).
 
 A fundamental difference between a MOLA and a teachable ML system  (Figure~\ref{fig:typemlsystems}.b) is that a MOLA directly and simultaneously interact with the whole community of experts, including people with different kinds of knowledge and experience, different opinions and approaches to the subject matter, and different degrees of commitment to the system. 
 
 A community of experts creates a whole set of new requirements and challenges for MOLAs related to the diversity of points-of-view, to the communal management of the knowledge artifacts, and to the importance of the personal relationships. By analogy to other open online communities such as \emph{Wikipedia}, we should expect that the complexities and conflicts of the community around a particular body of knowledge will impact the development and maintenance of the MOLA, but at the same time that the system will  greatly benefit from the strength of its community. 
 
However, unlike the wikipedia page which is concrete, clearly visible, and easily reconfigured by the members of the community, the knowledge a ML system possesses, which determines how it actually behaves, is much more complicated to represent, change, and assess. 
Moreover, one of the key assumptions of traditional ML systems is that the ground knowledge (often represented through labeled examples) is coherent and easily extracted from the experts.
 
 The assumption about coherence of knowledge is, in reality, not true for the majority of the knowledge areas and interesting problems or tasks.
 But instead of ignoring or hiding under the covers the plurality of opinions, as it is often done in many ML systems, MOLAs have no alternative but to be constructed to explicitly handle contradictory facts and judgments. At the same time, by acknowledging the complexities of real-world knowledge, a machine which learns from a community is more likely to truly acquire the knowledge of a group of experts.

As such, MOLAs are a great opportunity for domains where disagreement about knowledge is common, like in health, law, and science. Also, in domains where the actual knowledge is widely dispersed among community members, such as in technical support and customer care, a MOLA allows many individuals to contribute their piece of knowledge to an integrated and orchestrated intelligent system. Finally, since MOLAs must be able to learn from multiple people, continuously digesting and updating their knowledge, they are very appropriate for domains where knowledge varies through time, such as news, economy, sports, and politics.

\section{\uppercase{A Community-Created AI Assistant}}

To understand better the new requirements and challenges posed by MOLAs, let us explore a hypothetical example of how a professional community could work together to create a professional ML system which would support them in the everyday work. Suppose a community of nutritionists gathers together to create an AI assistant, in a MOLA, to support food choices of people with diabetes.
For that, let us forget for a moment the technical difficulties to make a MOLA actually work well, which we will address later in the paper. 

In this example a group of nutritionists, working together, decides they want to develop a conversational system which will dialogue with patients to support healthy food decisions. They decide to use a established conversational MOLA platform to which they input some samples of typical dialogues created by a sub-group of the community. To improve the AI assistant, they interact with it, concurrently, simulating patients, and as the system make mistakes they teach it the right answers by providing new concepts, guidelines, and information about foods. If the system detects that different nutritionists are providing information which appears to be contradictory, it poses back the issue to the community, which in a special forum discusses and tries to reach a conclusion. 

The work gets to a point where the community believes it is time to do a \emph{beta} release. Many of them volunteer patients as subjects for a first trial, and work together to monitor the system's conversations with the trial patients. The results are promising but the patients seem to be having a hard time understanding the physicians' language. They then decide to invite to their community a group of nurses who routinely explain dietary restrictions and options to diabetic patients. As the nurses work with the AI assistant, teaching it to talk using a simpler language, a sub-group of physicians keeps monitoring the quality of the advice to make sure that all information is medically correct.

After the successful first release, the community notices that the AI assistant is not being well received by some ethnic groups. They decide to create a task-force, spearheaded by physicians familiar with the diet of such groups, to enhance the AI assistant to address the cultural food particularities of that group. Similarly, some groups inside the community start to work on translating the assistant to other languages. As scientific knowledge about food and diabetes change, or new foods and recipes become available, members of the community update the knowledge of the AI assistant to handle those changes. As the community matures, being an influential member of the community becomes a mark of high professional status, similar to the notoriety some developers have today in open source communities.

It is not hard to see that such a system may be better than systems designed or built by a small group of individuals or organizations. In fact, there have been many attempts of creating AI-based nutritional assistants by start-ups and medical schools, at great effort and with limited impact. We have learned from the \emph{Linux} case and other open source software communities that tools and systems developed by organized communities are hard to beat, both in terms of scope and quality. We would expect that the community approach would warrant a similar level of success in this case of a nutritional AI assistant for diabetics.

\section{\uppercase{Challenges in Building MOLAs}}

In spite of the recent progress in machine teaching technology, discussed before, providing access to communities of experts to ML systems so they can build and improve them in a MOLA has several challenges. Let us list here some of them:
\begin{itemize}
    \item \textbf{Handling of contradictions:} learning from multiple teachers inevitably leads to receiving contradictory information. This problem already exists in today's ML training methods but it is worsened by the natural diversity of a community of experts and by their use of high-level teaching modalities.
    \item \textbf{Simultaneity:} a MOLA must learn from multiple teachers at the same time, what imposes additional burdens in terms of consistency and coherency. Also, teachers may experience disruptions in system behavior from one use to the next since the system may have changed due to interaction with experts with different points of view.
    \item \textbf{Levels of expertise:} a community of experts is likely to congregate members with very different degrees of expertise, proficiency, and even domain language familiarity. Orchestrating the contributions from teachers of diverse expertise levels is likely to require a multiplicity of levels of knowledge representation, and personalized weighting of the certainty about what is taught.
    \item \textbf{Evaluation consistency:} in a community it is likely to be disagreements about what a good performance is, both in terms of behavior and knowledge. Managing evaluations so they are coherent is a key problem to be dealt by MOLAs.
\end{itemize}

Notice that most of the challenges are, in fact, dual challenges: a technical one, related to ML algorithms able to handle some of those issues; and a user experience and interface design one, related to determining interface actions and modes which best enable the teacher to convey knowledge to the AI.

\section{\uppercase{When a Community Makes a Difference}}

The history of open source software and open online projects such as \emph{Wikipedia} and \emph{Bitcoin} have shown that communities have capabilities and emerging properties which individuals and even large organizations do not have. ML systems built with MOLAs can benefit from such capabilities to simplify their development process and to improve the quality of the final system in ways that would be hard to achieve with either developers or a small group of domain experts. We list now some of those capabilities:
\begin{itemize}
    \item \textbf{Consensus on the real needs:} a diverse group of experts is often better to arrive to a conclusion about what a good system must do and how to achieve it, as seen in open source software.
    \item \textbf{Motivation and resiliency:} a common problem in ML projects is that experts become unavailable after a period of time, and that developers come and go. However, open source projects are remarkably resilient to changes in the community, environment, and personnel.
    \item \textbf{Group understanding of the machine:} understanding what is happening to a ML system (especially in the case of complex errors) may be best achieved by a community effort which brings together people with different talents, as we have seen, for instance, in the forking decisions by crypto-currency communities.
    \item \textbf{High quality and reliability:} community-based development has a history of producing high quality and resilient products (for example, as shown by \emph{Linux}) because it congregates multiple points-of-view and skills.
    \item \textbf{Improved data gathering:} communities can create task-forces to tackle hard, undesirable, or time-consuming tasks. In particular, a large community of domain experts is likely to be more successful in gathering large amounts and more diverse data than a small group of experts.
    \item \textbf{Tool building:} in a communal spirit it is easier to gather groups of people with the skills, time, and motivation needed to create tools which may speed up development and maintenance. A great example is the emergence of \emph{miners} in Bitcoin.
    \item \textbf{Inclusiveness of knowledge:} learning from a community of experts is likely to assure that most of the different views about a given domain are represented and captured.
\end{itemize}

Of course the MOLA path is not free from risks and possible negative societal impacts. As we have seen in Wikipedia, sub-groups of a community can organize themselves to hijack content and push their own, often sectarian agendas. Similarly, we have seen attacks to consensus-based open structures such as in the case of \emph{Ethereum}. There is also the important issue of to whom belongs the authorship of a MOLA, since it is an aggregation of knowledge from multiple individuals (see \cite{blackwell2015}). As in any ML system, bias and hate can also creep in, not only from data but also from particular sub-groups of the community, as we saw with \emph{Tay}~\cite{neff2016}. But like in the case of open source software, working within an open structure also provides a good foundation to mitigate many of those issues and to fight against their negative consequences.














\section{\uppercase{Final Discussion}}

We believe that as machine teaching progresses and opens up the construction of ML systems to domain experts, the perspective of more intelligent and complex ML systems increases. Concurrently, the transition from training to teaching may make the development process more humane. And as it becomes easier to establish open community efforts to build ML systems, we can also expect them to be more inclusive in terms of diversity of knowledge, and more comprehensive results and adoption. In particular, MOLAs may be key to unlock the deployment of intelligent systems in complex, diverse domains such as health, law, and economics.

However, to go from the trainable DL systems of today to teachable ML systems which can handle the complexity and diversity of a community-based effort requires addressing the many challenges we discussed. There is a lot of exciting research work to be done, both on the underlining engines of machine learning, on the interface and user experience of the teachers, and on the organization and management of such communal efforts in this new context.


We, in the information technology disciplines, have been very lucky to experience and benefit from open community work from the early days of the \emph{Ethernet}, renewed in construction of the \emph{Internet}, reinvigorated by the \emph{Linux} movement, and reflected today in the plethora of open source code and data available. MOLAs may be the key to bring similar community efforts to the practice of almost every domain of knowledge. In fact, we may be struggling to build reliable large-scale AI systems simply because we have narrowed down to ML experts the ability to build them. It is time to open up ML to non-AI experts and their communities, and possibly discover that by doing so it becomes a lot simpler to create and maintain intelligent systems.

It may take a village to raise an AI system.




\bibliographystyle{apalike}
\bibliography{references}

\begin{thebibliography}{}

\bibitem[Amershi et~al., 2014]{amershi2014}
Amershi, S., Cakmak, M., Knox, W.~B., and Kulesza, T. (2014).
\newblock Power to the people: The role of humans in interactive machine
  learning.
\newblock {\em AI Magazine}, 35(4):105--120.

\bibitem[Amershi et~al., 2015]{amershietal2015}
Amershi, S., Chickering, M., Drucker, S., et~al. (2015).
\newblock Modeltracker: Redesigning performance analysis tools for machine
  learning.
\newblock In {\em Proc. of CHI'15}, pages 337--346. ACM.

\bibitem[Aroyo and Welty, 2015]{aroyowelty2015}
Aroyo, L. and Welty, C. (2015).
\newblock Truth is a lie: Crowd truth and the seven myths of human annotation.
\newblock {\em AI Magazine}, 36(1):15--24.

\bibitem[Azaria et~al., 2016]{azariaetal2016}
Azaria, A., Krishnamurthy, J., and Mitchell, T.~M. (2016).
\newblock Instructable intelligent personal agent.
\newblock In {\em Proc. of AAAI'16}.

\bibitem[Blackwell, 2015]{blackwell2015}
Blackwell, A.~F. (2015).
\newblock Interacting with an inferred world: the challenge of machine learning
  for humane computer interaction.
\newblock In {\em Proc. of The Fifth Decennial Aarhus Conference on Critical
  Alternatives}, pages 169--180. Aarhus University Press.

\bibitem[Darwiche, 2018]{darwiche2018}
Darwiche, A. (2018).
\newblock Human-level intelligence or animal-like abilities?
\newblock {\em CACM}, 61:56--67.

\bibitem[Dudley and Kristensson, 2018]{dudleykristensson2018}
Dudley, J.~J. and Kristensson, P.~O. (2018).
\newblock A review of user interface design for interactive machine learning.
\newblock {\em ACM Transactions on Interactive Intelligent Systems (TiiS)},
  8(2):8.

\bibitem[Fails and Olsen~Jr, 2003]{failsolsen2003}
Fails, J.~A. and Olsen~Jr, D.~R. (2003).
\newblock Interactive machine learning.
\newblock In {\em Proc. of IU'03}, pages 39--45. ACM.

\bibitem[Fan et~al., 2018]{fanetal2018}
Fan, Y., Tian, F., Qin, T., Li, X.-Y., and Liu, T.-Y. (2018).
\newblock Learning to teach.
\newblock In {\em Proc. of ICML'18}.

\bibitem[Fiebrink et~al., 2011]{fiebrinketal2011}
Fiebrink, R., Cook, P.~R., and Trueman, D. (2011).
\newblock Human model evaluation in interactive supervised learning.
\newblock In {\em Proc. of CHI'11}.

\bibitem[Flexer and Grill, 2016]{flexergrill2016}
Flexer, A. and Grill, T. (2016).
\newblock The problem of limited inter-rater agreement in modelling music
  similarity.
\newblock {\em Journal of New Music Research}, 45(3):239--251.

\bibitem[Fogarty et~al., 2008]{fogartyetal2008}
Fogarty, J., Tan, D., Kapoor, A., and Winder, S. (2008).
\newblock Cueflik: interactive concept learning in image search.
\newblock In {\em Proc. of CHI'08}, pages 29--38. ACM.

\bibitem[Gurari et~al., 2015]{gurarietal015}
Gurari, D., Theriault, D., Sameki, M., et~al. (2015).
\newblock How to collect segmentations for biomedical images? a benchmark
  evaluating the performance of experts, crowdsourced non-experts, and
  algorithms.
\newblock In {\em 2015 IEEE Winter Conference on Applications of Computer
  Vision}, pages 1169--1176.

\bibitem[Hancock et~al., 2018]{hancock2018}
Hancock, B., Varma, P., Wang, S., Bringmann, M., Liang, P., and R{\'e}, C.
  (2018).
\newblock Training classifiers with natural language explanations.
\newblock {\em arXiv preprint arXiv:1805.03818}.

\bibitem[Hood et~al., 2015]{hoodetal2015}
Hood, D., Lemaignan, S., and Dillenbourg, P. (2015).
\newblock When children teach a robot to write: An autonomous teachable
  humanoid which uses simulated handwriting.
\newblock In {\em Proc. of HRI'15}, pages 83--90.

\bibitem[Kaochar et~al., 2011]{kaocharetal2011}
Kaochar, T., Peralta, R.~T., Morrison, C.~T., Fasel, I.~R., Walsh, T.~J., and
  Cohen, P.~R. (2011).
\newblock Towards understanding how humans teach robots.
\newblock In {\em Proc. of UMAP'11}, pages 347--352.

\bibitem[Kulesza et~al., 2015]{kuleszaetal2015}
Kulesza, T., Burnett, M., Wong, W.-K., and Stumpf, S. (2015).
\newblock Principles of explanatory debugging to personalize interactive
  machine learning.
\newblock In {\em Proc. of IUI'15}, pages 126--137. ACM.

\bibitem[Leu et~al., 2017]{leu2017}
Leu, G., Lakshika, E., Tang, J., Merrick, K., and Barlow, M. (2017).
\newblock Machine education-the way forward for achieving trust-enabled machine
  agents.
\newblock In {\em Proc. of the NIPS'17 Workshop on Teaching Machines, Robots,
  and Humans}.

\bibitem[Liu et~al., 2017]{liuetal2017}
Liu, W., Dai, B., Humayun, A., Tay, C., Yu, C., Smith, L.~B., Rehg, J.~M., and
  Song, L. (2017).
\newblock Iterative machine teaching.
\newblock In {\em Proc. of ICML'17}, pages 2149--2158.

\bibitem[Mitchell et~al., 2015]{mitchell2015}
Mitchell, T., Cohen, W., Hruschka, E., et~al. (2015).
\newblock Never-ending learning.
\newblock In {\em Proc. of AAAI'15}.

\bibitem[Neff and Nagy, 2016]{neff2016}
Neff, G. and Nagy, P. (2016).
\newblock Automation, algorithms, and politics | talking to bots: symbiotic
  agency and the case of tay.
\newblock {\em International Journal of Communication}, 10:17.

\bibitem[Quintana-Amate et~al., 2015]{quintanaetal2015}
Quintana-Amate, S., Bermell-Garcia, P., and Tiwari, A. (2015).
\newblock Transforming expertise into knowledge-based engineering tools: A
  survey of knowledge sourcing in the context of engineering design.
\newblock {\em Knowledge-Based Systems}, 84:89--97.

\bibitem[Roy et~al., 2018]{royetal2018}
Roy, S., Kieson, E., Abramson, C., and Crick, C. (2018).
\newblock Using human reinforcement learning models to improve robots as
  teachers.
\newblock In {\em Proc. of HRI'18}, pages 225--226. ACM.

\bibitem[Sarkar, 2015]{sarkar2015}
Sarkar, A. (2015).
\newblock Confidence, command, complexity: metamodels for structured
  interaction with machine intelligence.
\newblock In {\em Proc. of the PPIG'15}.

\bibitem[Senft et~al., 2017]{senftetal2017}
Senft, E., Lemaignan, S., Baxter, P.~E., and Belpaeme, T. (2017).
\newblock Leveraging human inputs in interactive machine learning for human
  robot interaction.
\newblock In {\em Proc. of HRI'17}, pages 281--282.

\bibitem[Simard et~al., 2017]{simard2017}
Simard, P.~Y., Amershi, S., Chickering, D.~M., et~al. (2017).
\newblock Machine teaching: A new paradigm for building machine learning
  systems.
\newblock {\em arXiv preprint arXiv:1707.06742}.

\bibitem[Stumpf et~al., 2009]{stumpf2009}
Stumpf, S., Rajaram, V., Li, L., et~al. (2009).
\newblock Interacting meaningfully with machine learning systems: Three
  experiments.
\newblock {\em International Journal of Human-Computer Studies},
  67(8):639--662.

\bibitem[Tecuci et~al., 2016]{tecucietal2016}
Tecuci, G., Marcu, D., Boicu, M., and Schum, D.~A. (2016).
\newblock {\em Knowledge Engineering: Building Cognitive Assistants for
  Evidence-based Reasoning}.
\newblock Cambridge University Press.

\bibitem[Thomaz and Breazeal, 2008]{thomazbreazeal2008}
Thomaz, A.~L. and Breazeal, C. (2008).
\newblock Teachable robots: Understanding human teaching behavior to build more
  effective robot learners.
\newblock {\em Artificial Intelligence}, 172(6-7):716--737.

\bibitem[Zhu, 2015]{zhu2015}
Zhu, X. (2015).
\newblock Machine teaching: An inverse problem to machine learning and an
  approach toward optimal education.
\newblock In {\em Proc. of AAAI'15}.

\end{thebibliography}

\end{document}